\documentclass[10pt, openany]{book}

\usepackage[]{graphicx}
\usepackage{booktabs}
\usepackage[]{color}
\usepackage{alltt}
\usepackage[T1]{fontenc}
\usepackage{adjustbox}
\usepackage[utf8]{inputenc}
\usepackage{threeparttable}
\usepackage[top=100pt,bottom=100pt,left=68pt,right=66pt]{geometry}

\usepackage{lipsum}

\usepackage[swedish, english]{babel}

\usepackage{fancyhdr}
\usepackage{titlesec}
\titleformat{\chapter}
   {\normalfont\LARGE\bfseries}{\thechapter.}{1em}{}
\titlespacing{\chapter}{0pt}{50pt}{2\baselineskip}

\usepackage{float}
\floatstyle{plaintop}
\restylefloat{table}

\usepackage[tableposition=top]{caption}

\usepackage{hyperref}
\hypersetup{hidelinks,linkcolor = black} 

\graphicspath{ {images/} }

\frontmatter

\begin{document}


\begin{titlepage}
	\clearpage\thispagestyle{empty}
	\centering
	\vspace{2cm}

	{\large Computer Networks - COEN 233 \par}
	\vspace{4cm}
	{\Huge \textbf{QUIC - A Quick Study}} \\
	\vspace{1cm}
	{\large \textbf{Quick UDP Internet Connections} \par}
	\vspace{4cm}
	{\normalsize Puneet Kumar \\ 
	             SCU-ID: 00001424550 \par}
	\vspace{2cm}

    \includegraphics[scale=0.75]{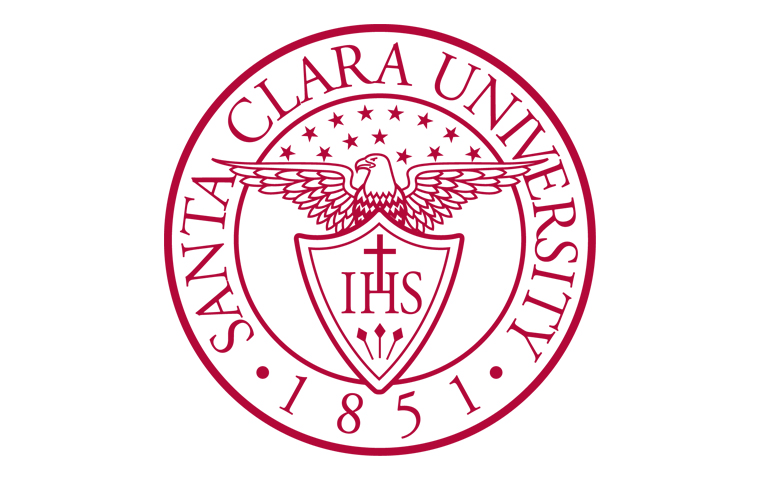}
    
    \vspace{2cm}
    
	{\normalsize Santa Clara University \\ 
		Computer Networks: COEN-233 \\
		Department of Computer Science \par}
		
	{\normalsize California, USA \par}
	\vspace{2cm}
	
	\pagebreak

\end{titlepage}

\chapter{Audience}
\label{chapter_1}
\large This document is intended to explain the new transport protocol which is going to revolutionize the network industry. It is important to have prior knowledge of the transport protocols (TCP and UDP). QUIC is exploring it's way into TCP world on a fast pace. This documents explains the protocol in a great detail. It targets audience of the class who are taking taken COEN - 233 (Computer Networks), as fundamentals to understand this document was explained in the class, as in lecture slides, OSI layer, transport layer protocols and it's interaction with layer 3(Network layer) and layer 7 (Application layer) are covered.

\tableofcontents{}
\listoffigures
\listoftables

\mainmatter

\chapter{\large Introduction}
\label{intro}
\large Main responsibility of a transport protocol is to support communication between two end-to-end entities. These entities can be hosts or devices, such as routers, firewalls etc. Transport protocol provides mechanism to have a virtually circuited route between these two end-to-end devices. Transport protocol are of two types: Connection oriented and connection-less. As these name suggest, connection oriented requires additional work of setting up connections and provide reliability in terms of packet re-ordering, congestion control, reliable delivery of packets. On the other hand connection-less transport protocol adapts, sends and forgets mechanism. It's entire purpose is to deliver the datagram without having to worry about it's delivery. There are two major protocols, TCP (Transmission Control Protocol) \cite{postel1981transmission} and UDP (User Datagram Protocol) \cite{postel1980user}, used for connection oriented and connection-less services respectively. Being a connection oriented protocol, TCP provides a reliable end-to-end connection and have congestion control mechanisms to avoid buffer overflow at the receiver side.
\begin{figure}[H]
    \centering
    \includegraphics[width=0.4 \linewidth]{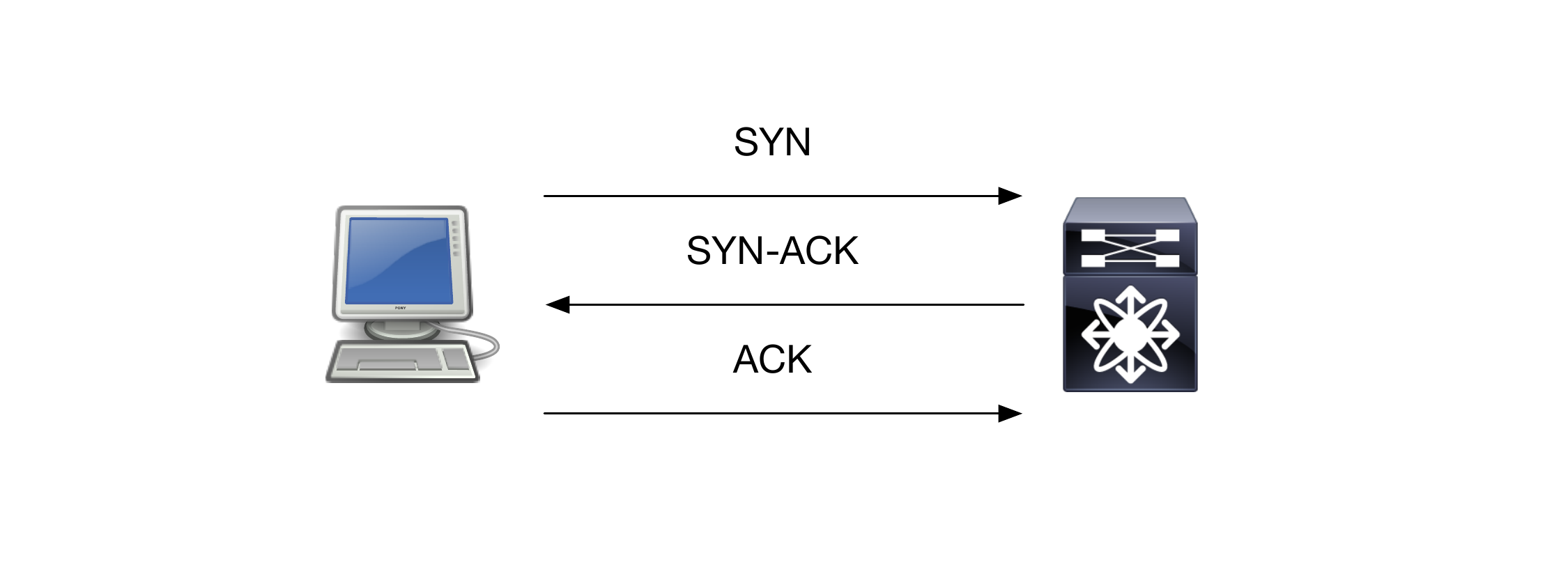}
    \caption{TCP Connection Establishment}
    \label{fig:tcp_conn}
\end{figure}

\begin{table*}[htp]
\begin{center}
\caption{TCP Versions Comparison}
\resizebox{\columnwidth}{!}{%
\begin{tabular}{@{}ccccc@{}}
\toprule
Version      & Feedback          & Required Changes         & Benefits                                                                     & Fairness      \\ \midrule
Reno (New)   & Loss              & --                       & --                                                                           & Delay         \\
Vegas        & Delay             & Sender                   & Less Loss                                                                    & Propotional   \\
High Speed   & Loss              & Sender                   & High Bandwidth                                                               &               \\
BIC          & Loss              & Sender                   & High Bandwidth                                                               &               \\
CUBIC        & Loss              & Sender                   & High Bandwidth                                                               &               \\
C2TCP        & Loss/Delay        & Sender                   & \begin{tabular}[c]{@{}c@{}}Ultra-low latency \\ high bandwidth\end{tabular}  &               \\
NATCP        & Multi-bit Signal  & Sender                   & \begin{tabular}[c]{@{}c@{}}Near Optimal \\ Performance\end{tabular}          &               \\
Elastic-TCP  & Loss and Delay    & Sender                   & \begin{tabular}[c]{@{}c@{}}High Bandwidth/Short\\ Long-distance\end{tabular} &               \\
Agile-TCP    & Loss              & Sender                   & \begin{tabular}[c]{@{}c@{}}High Bandwidth/Short \\ Distance\end{tabular}     &               \\
H-TCP        & Loss              & Sender                   & High Bandwidth                                                               &               \\
FAST         & Delay             & Sender                   & High Bandwidth                                                               & Proportional  \\
Compound TCP & Loss/Delay        & Sender                   & High Bandwidth                                                               & Proportional  \\
Westwood     & Loss/Delay        & Sender                   & L                                                                            &               \\
Jersey       & Loss/Delay        & Sender                   & L                                                                            &               \\
BBR          & Delay             & Sender                   & BLVC, Bufferbloat                                                            &               \\
CLAMP        & Multi-bit Signal  & Receiver, Router         & V                                                                            & Max-Min       \\
TFRC         & Loss              & Sender, Receiver         & No Retransmission                                                            & Minimum Delay \\
XCP          & Multi-bit Signal  & Sender, Receiver, Router & BLFC                                                                         & Max-Min       \\
VCP          & 2-bit Signal      & Sender, Receiver, Router & BLF                                                                          & Proportional  \\
MaxNet       & Multi-bit Signal  & Sender, Receiver, Router & BLFSC                                                                        & Max-Min       \\
JetMax       & Multi-bit Signal  & Sender, Receiver, Router & High Bandwidth                                                               & Max-Min       \\
RED          & Loss              & Router                   & Reduced Delay                                                                &               \\
ECN          & Single-bit Signal & Sender, Receiver, Router & Reduced Loss                                    \label{tcp_comparison_all}                             &               \\ \bottomrule
\end{tabular}
}
\end{center}

\end{table*}

\section{\large Connection Overhead}
\large Since TCP is the most important connection oriented protocol, it has been a center point of the research. Several improved version of TCP have been published in the last couple of decades(refer Table \ref{tcp_comparison_all}). These TCP versions have mainly focused on the throughput \cite{floyd2004newreno, xu2004binary}. Albeit, throughput is the main factor for determining the quality of traffic, TCP might not be for suitable for a lot of different types of traffic. Some of the examples of such traffic are: IoT (Internet of Things) \cite{xia2012internet}, IIoT (Industrial Internet of Things) \cite{wurm2016security}. Traffic in these technologies are generally characterized by short-lived bursts of exchanging small data chunks. For such short lived connections, establishing connections every single time before an exchange of short live burst data is a lot of overhead. Connection start up latency in TCP is high, as it requires two round trips (RTT) for declaring a peer-to-peer connection alive and ready.

\large Generally, applications which are required to have connection oriented transport protocol such as HTTPS \cite{rescorla2000rfc}, have some form of security enabled on top of transport layer. One of the most popular protocol used for such purpose is TLS (Transport Layer Security) \cite{dierks2008rfc}. This protocol provides security over the TCP by encrypting everything in application layer. Encryption is based on the keys exchange happened between end entities. This process adds one or two RTTs into the additional two RTTs in connection establishment between end points after connection establishment. Securing transport layer is necessary as it provides symmetric encryption which enables authenticated, confidential and integrity-preserved communication between two devices. TLS protocol has it's own handshake like TCP and it is unique per connection. So total 3-4 RTTs are required to establish and secure a connection oriented connection between end-to-end hosts. 

\large As we have seen, it takes 3-4 RTT just to establish a TCP connection. Networks which are lossy in nature, such as wireless networks, will pay a heavy price of connection overhead. Everytime an IoT device gets disconnected from the network, to connect to the desired  device again, TCP connection needs to be re-established. If a network has a lot of lossy connections, the amount of connection establishment overhead can be enormous. 

\begin{figure}[H]
    \centering
    \includegraphics[width=0.3 \linewidth]{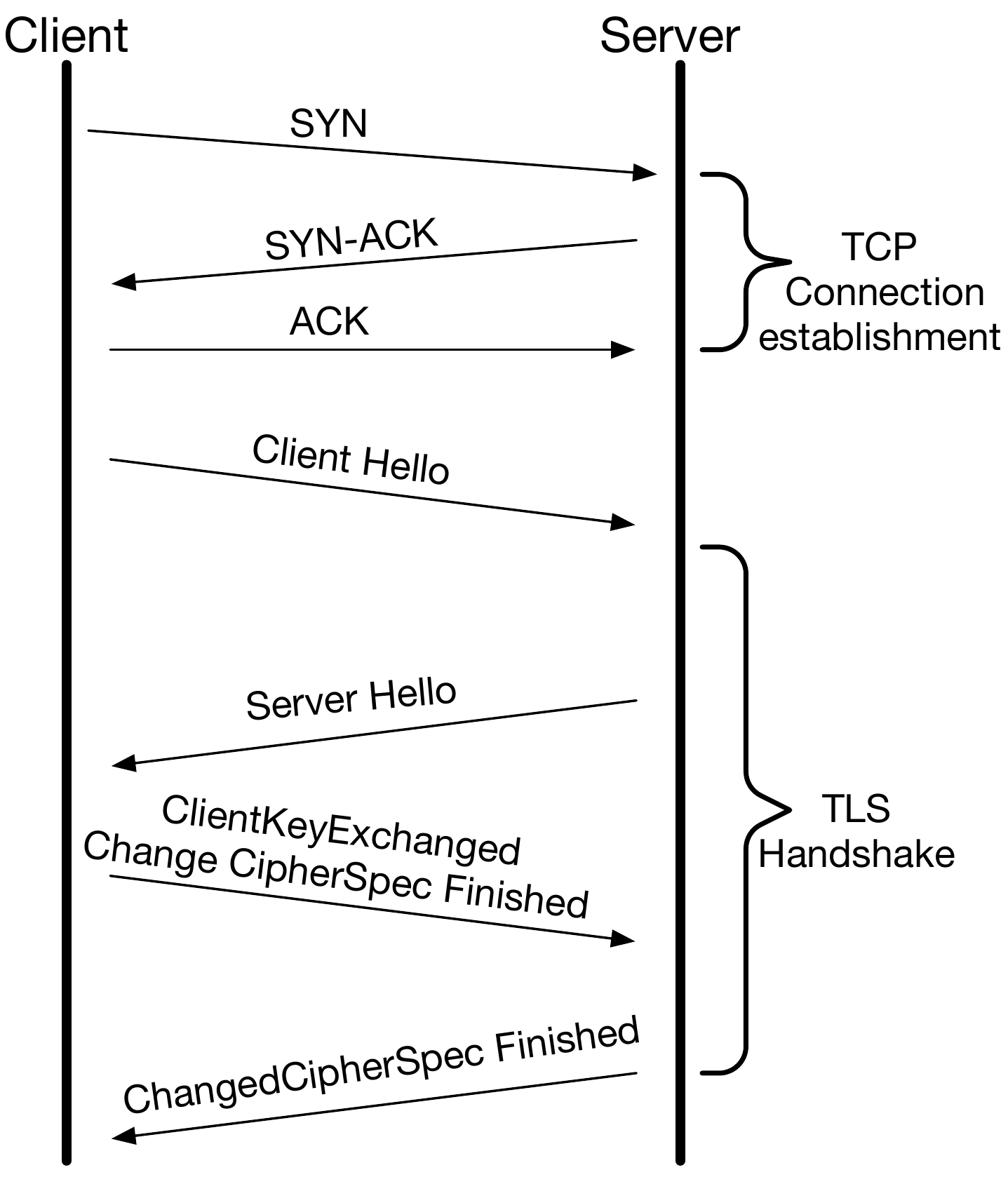}
    \caption{Total number of RTTs}
    \label{fig:total_rtt}
\end{figure}
\section{\large Migrating Flows}
\label{tcp_con_mig}
\large Limitations of TCP is not stopped only to connection overhead. Another major price TCP connection pays is flow migration. Flow is generally identified by a combination of several factors to identify single connection. Parameters involved in identifying a flow are source IP address, destination IP address, source port number, destination port number and the protocol (called 5-tuple). Some of these parameters (port numbers) are unique for a connection while others are common across device (ip address). Changing the application or device can lead to a different identifier than the previous one. This will eventually lead to tear down the establish TCP connection. Such condition will force devices to establish a new TCP connection and perhaps negotiate TLS handshake again. This could lead to a catastrophic event, imagine a user is paying a credit card bill and the money has been reduced from his account, but while contacting the vendor, connection is teared down. All session information prior to this connection is gone and now in new connection, packets have no information about prior bill pay. Second limitation arises due to flow migration is NAT rebinding. Generally, users are behind firewall in their LAN network, with non-routable IP (such as 192.168.x.x or 172.16.x.x), these IP and port makes a NAT table in the outgoing router to identify each device. Upon connection migration, this table can be inconsistent and can lead to breach in LAN hosts securities.

\section{\large Multiple Connections to Same Server}
\label{multiplexing}
\large Some applications, specially the web applications, such as HTTP \cite{network1999rfc} require to have multiple TCP connections established with the server in order to fetch data. These connection occupies bandwidth and bring connection overhead with them. Currently there is only one flavor of TCP out there called SPDY \cite{belshe2008draft} which supports multiplexing (SCTP is not a flavor of TCP). Problem with SPDY is that it is not supported by many networks and it's limited usage to HTTP2 hindered it's popularity.

\section{\large TCP Half-Open Connection}
\large Another limitation of TCP is half open connections. Receiving data in TCP is passive. It means that dropped connections are only detected by the sender and receiver has no way of detecting them. In order to understand this problem, lets take an example. Suppose a connection is being established between client and server. If a server is failed to receive an \texttt{ACK} from client, this port is continued to be occupied and resources will only be released once the TCP state transitions to \texttt{Closed State} after timer expires. 

\begin{figure}[H]
    \centering
    \includegraphics[width=0.3 \linewidth]{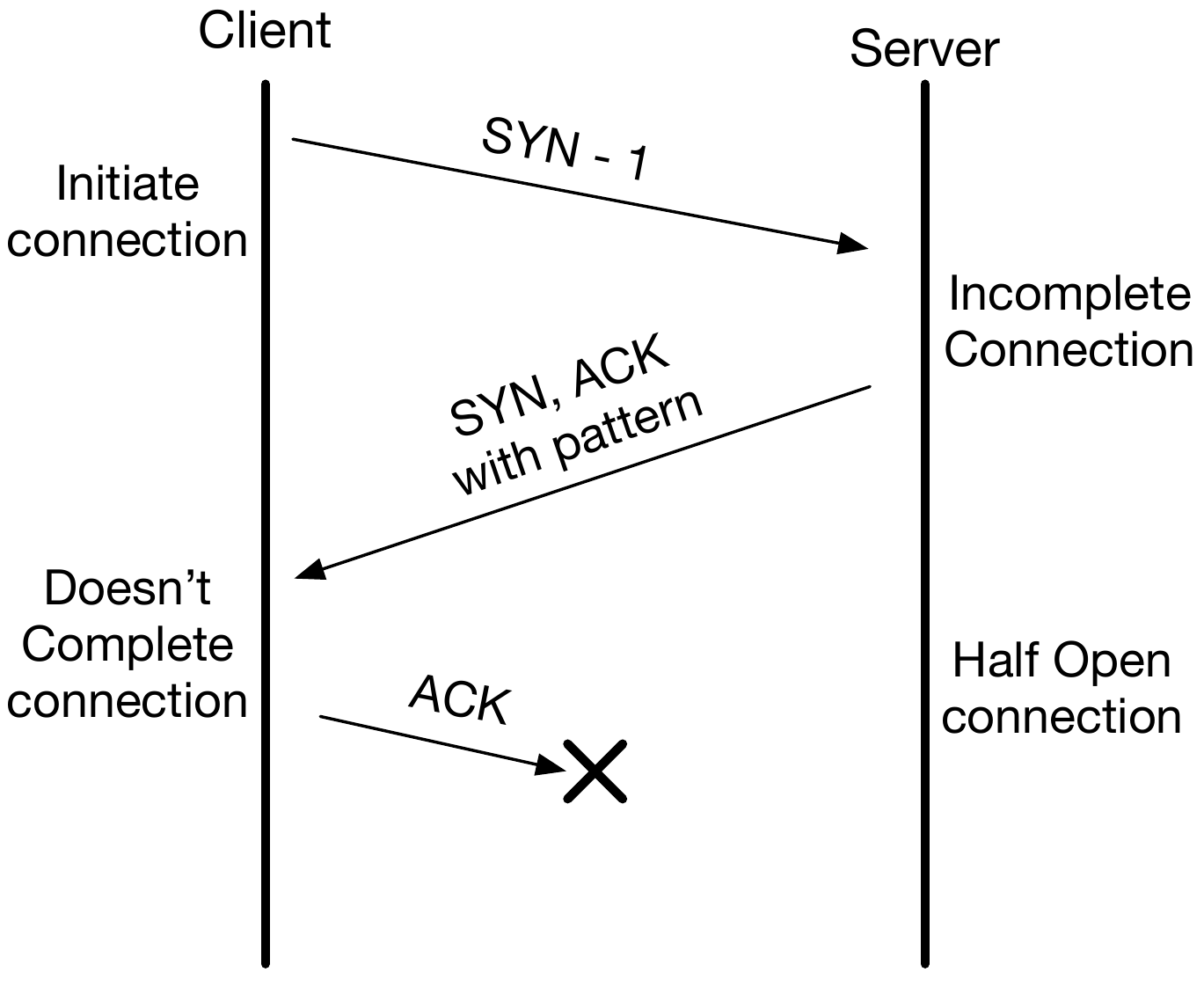}
    \caption{TCP Half-Open Connection}
    \label{fig:tcp_half_open}
\end{figure}
\section{\large Head-of-Line Blocking (HOL)}
\begin{figure}[H]
    \centering
    \includegraphics[width=0.7 \linewidth]{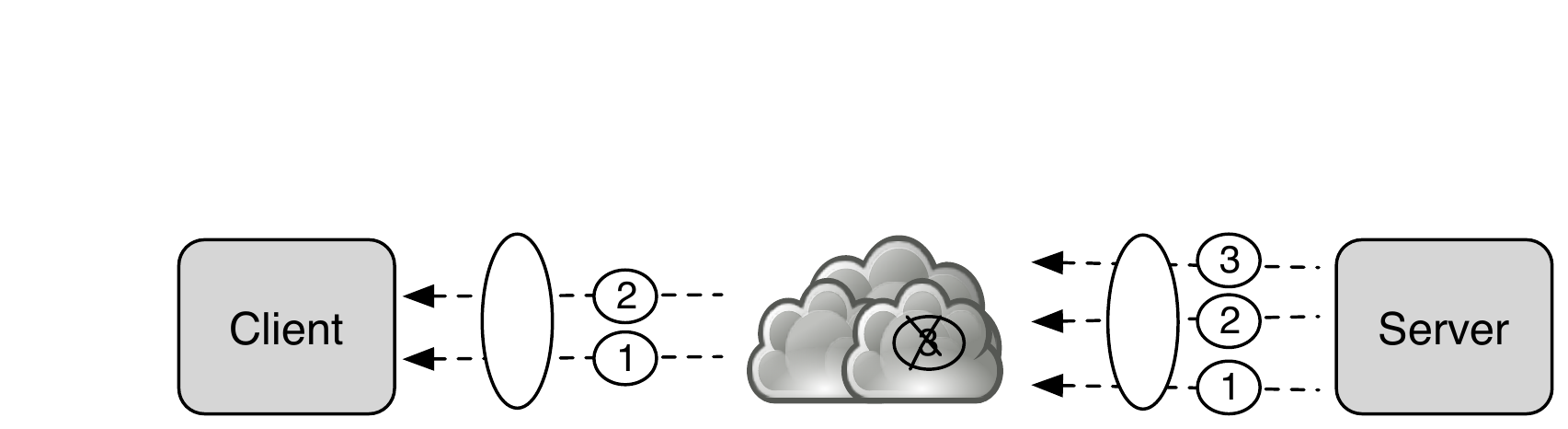}
    \caption{Head-of-Line Blocking}
    \label{fig:HeadOfLine}
\end{figure}

Head-of-Line blocking problem is a limitation phenomena, which occurs when resources are being held up by some entity in order to complete an action. As shown in figure \ref{fig:HeadOfLine}, three packets are coming from server and if packet 3 is dropped for some reason then packet 1 and 2 will not be pushed to application, until packet 3 is re-transmitted and packet re-ordering is completed. As one might guess, this phenomena only occurs if a connection oriented transport protocol is used. This out of delivery of packets can happen due to several reasons, such as lossy nature of the network, different paths taken by the packets and being dropped. HOL can significantly increase the packet reordering delay and which lead to consume resources and latency in packets being processed by network stack and application.

\section{\large Contribution}
With the increasing number of devices connected to internet, a new transport mechanism is needed to replace TCP to be able to address all the problems mentioned above. In this research report we will investigate and dissect QUIC \cite{kumar2019implementation} protocol. QUIC (Quick UDP Internet Connections) addresses these issues and bring a new perspective to the transport layer. Not only it provides fast transport but in terms of security it integrates TLS-1.3 which is the next generation transport layer security protocol. Chapter 2 will be focused on QUIC protocol specifically, chapter 3 will mainly talk about conclusion and explanation about the future aspects of the protocol.

\chapter{\large QUIC Protocol}
QUIC stands for Quick UDP Internet connections \cite{kumar2019implementation} and developed by Google. Initially this protocol was designed and developed for HTTP but later it was declared as general purpose transport layer protocol. This protocol provides security and reliability along with reduced connection and transport latency. Google has widely deployed QUIC in their servers and is currently in use. 

\begin{figure}[H]
    \centering
    \includegraphics[width=0.6 \linewidth]{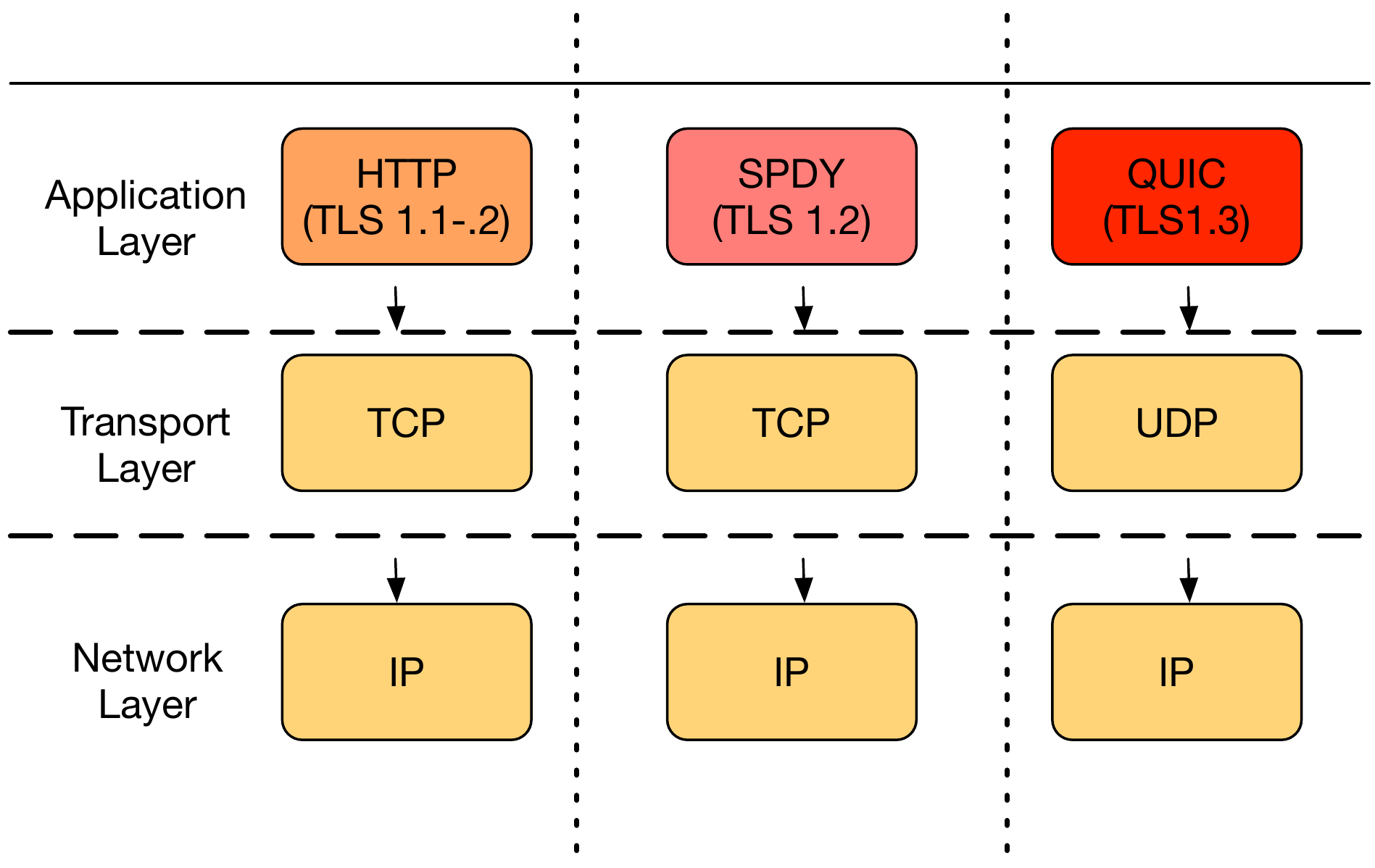}
    \caption{QUIC Protocol in Network Stack}
    \label{fig:quic_stack}
\end{figure}

As shown in figure \ref{fig:quic_stack}, QUIC sits in application layer unlike other traditional transport layer protocol such as TCP or UDP. QUIC runs on UDP to be compatible with all the middleboxes. When middleboxes encounter QUIC traffic, they see only UDP traffic. The main idea behind QUIC is to overcome TCP limitations discussed in Chapter \ref{intro}. As one might guess that the main transport protocol for QUIC, which is UDP, it is unreliable and doesn't provide any features needed for reliable connection.  

QUIC solves the problems mentioned in Chapter \ref{intro}. QUIC is very similar to TCP+TLS+HTTP2, but since it is implemented on top of UDP, all the connection reliability features are in application layer. As a self contained protocol, there are limitless opportunity for innovation in QUIC, which are not going to be possible with existing protocols as they bound to follow legacy. 

\section{\large QUIC Features}
In this section, we will elaborate about the shortcomings we mentioned in Chapter \ref{intro}.  If we compare QUIC with TCP+TLS+HTTP2. It brings the following advantages:

\begin{itemize}
    \item Connection establishment latency.
    \item Improved Congestion Control
    \item Multiplexing without head-of-line blocking
    \item Forward Error Correction
    \item Connection Migration
\end{itemize}

\begin{figure}[H]
    \centering
    \includegraphics[width=0.6 \linewidth]{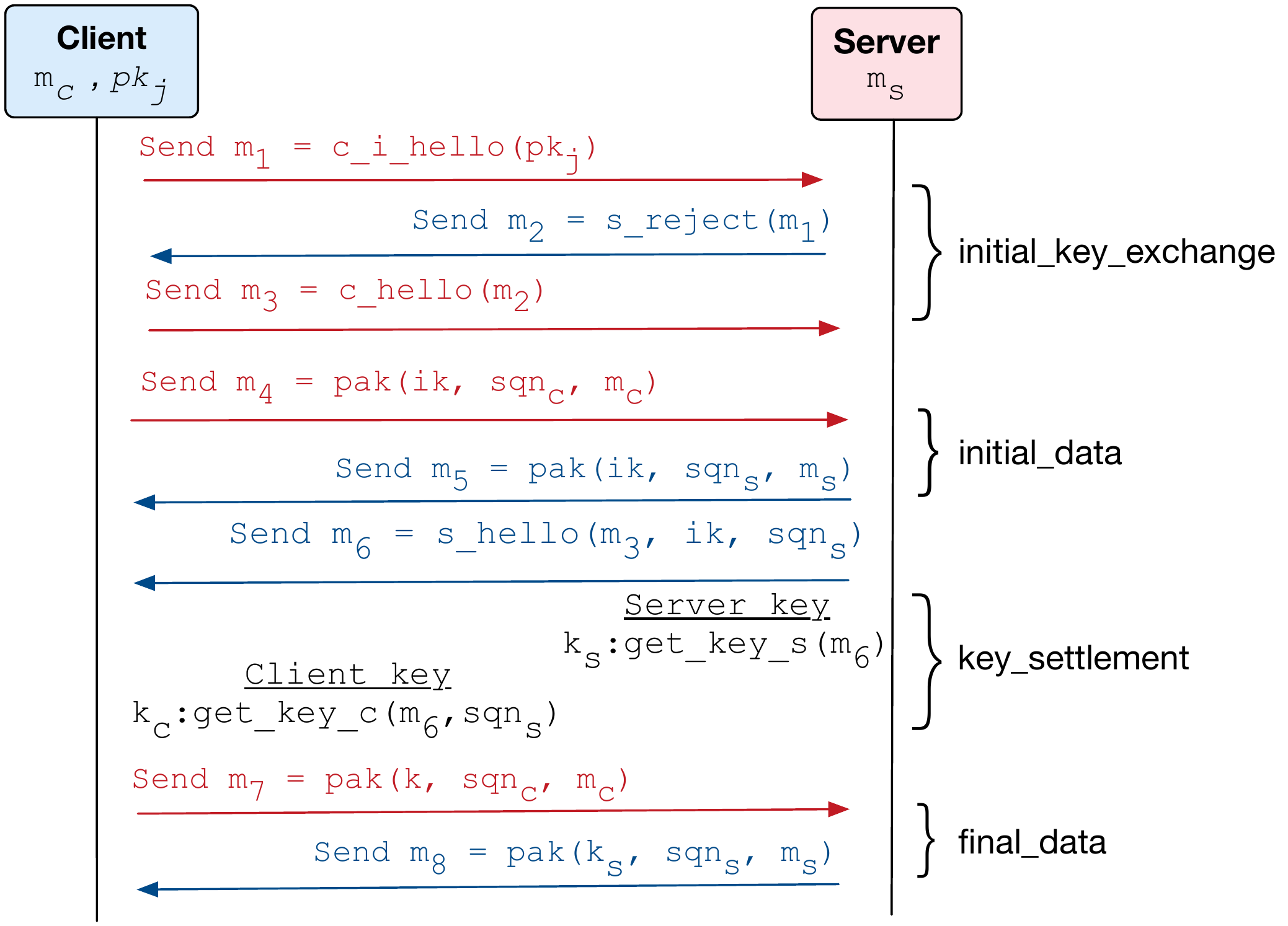}
    \caption{1 RTT QUIC connection establishment}
    \label{fig:1RTT}
\end{figure}

\subsection{\large Connection Establishment}
As mentioned in chapter \ref{intro}, TCP in conjunction with TLS, requires 3-4 RTTs to establish connection before data being encrypted and sent. This overhead makes applications slower and with the increasing demands of bandwidth and application availability, this overhead can be a bottleneck. QUIC solves this problem with having at most 1 RTT for fresh connection and 0 RTT for repetitive connections. If a connection is being established between client and server for the first time, then it must perform 1 RTT handshake in order to get the necessary information. This is done by client sending an empty \texttt{Client Hello (CHLO)} packet to server. Upon receiving a \texttt{CHLO} packet, server immediately send a \texttt{Rejection (REJ)} message to client. \texttt{REJ} contains information such as source address token, server's certificates etc. In order to move further in connection establishment, this information is necessary. Next time, the client sends a \texttt{CHLO}  packet, which can use the cache credentials from the previous connection to immediately send encrypted requests to the server.
\begin{figure}[H]
    \centering
    \includegraphics[width=0.6 \linewidth]{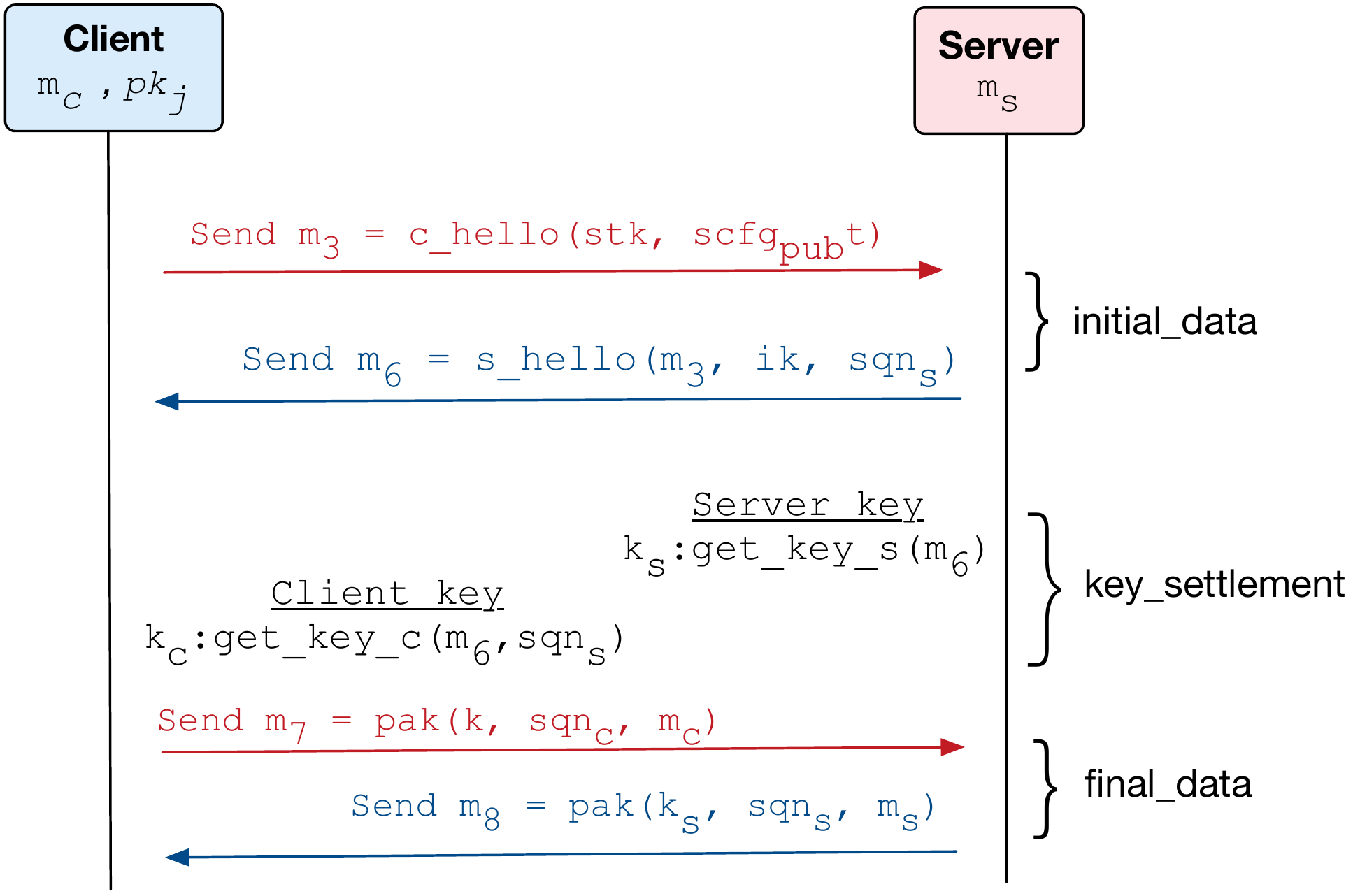}
    \caption{1 RTT QUIC connection establishment}
    \label{fig:0RTT}
\end{figure}

Figure \ref{fig:1RTT} shows the packet exchanges in QUIC connection establishment in 1-RTT scenario. 
In 0-RTT scenario, server and client have communicated in the past and no need to establish connection again. As shown in figure \ref{fig:0RTT}, first packet itself carry the data, there is no need for connection establishment or key exchanges. 

\subsection{\large Improved Congestion Control}
QUIC inherits TCP cubic algorithm for congestion control \cite{ha2008cubic}. Unlike TCP, QUIC has pluggable congestion control, and provides richer information to the congestion control algorithm. For instance, each packet whether it is original or re-transmitted, has a unique sequence number. This simple hack provides QUIC sender to distinguish between original and re-transmitted packets and solves TCP ambiguity problem \cite{gurtov2004resolving}. In addition to that QUIC ACKs also explicitly carry the delay between the acknowledgement sent and receipt of the packet. These sequence numbers are monotonically increasing numbers and do not get repeated in the connection lifetime.

\subsection{\large Multiplexing}
As mentioned in subsection \ref{multiplexing}, a client opens up multiple TCP connections to server to fetch data from server. QUIC solves this problem by having multiple streams over one UDP pipeline.
\begin{figure}[H]
    \centering
    \includegraphics[width=0.5 \linewidth]{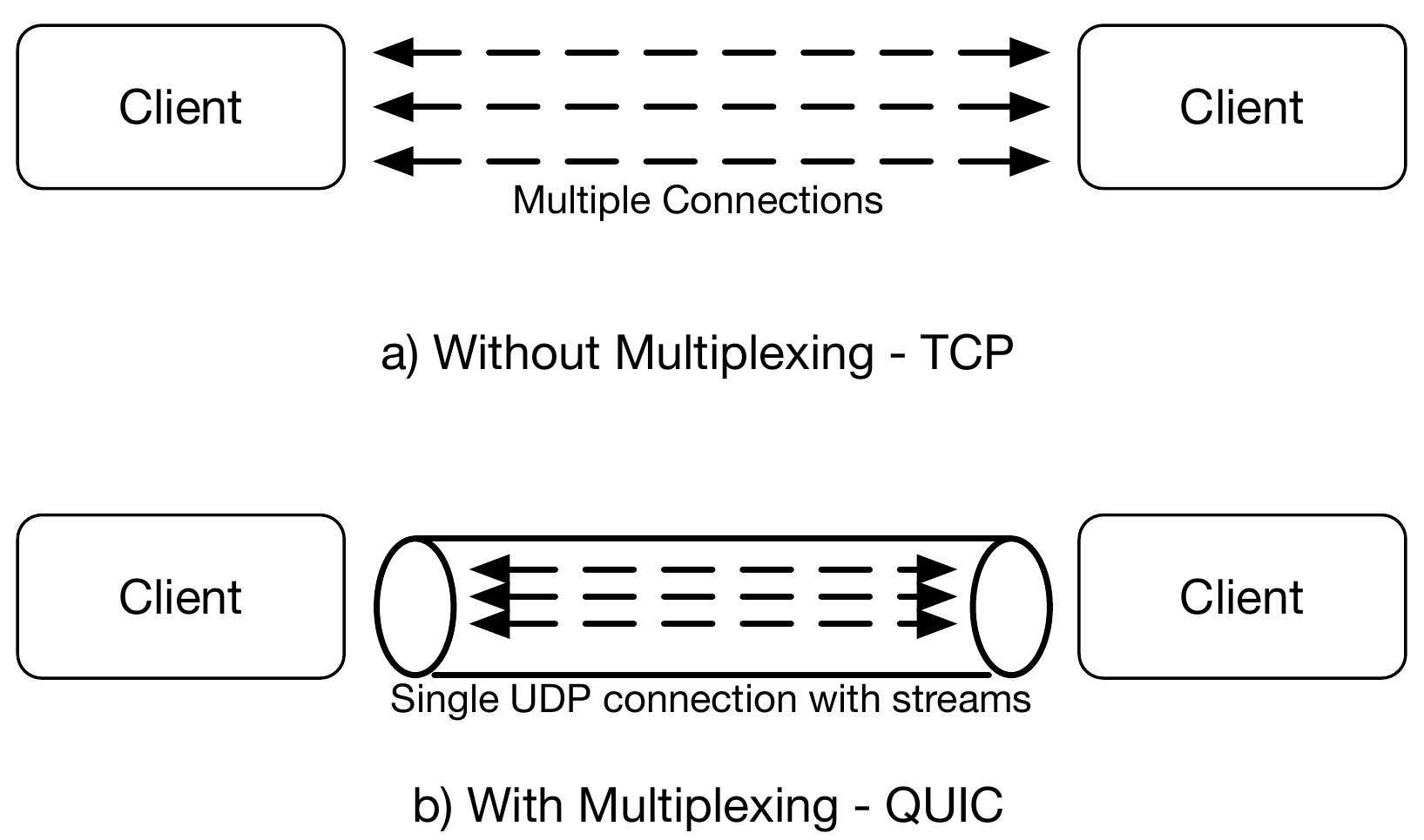}
    \caption{TCP vs QUIC Connections}
    \label{fig:multiplexing}
\end{figure}
In QUIC, there is only one UDP connection for transport, while in the usage of TCP, there are multiple connections being established.

\subsection{\large Forward Error Correction}
Forward error correction mechanism is to recover the lost packets, without having to worry about re-transmission. QUIC can complement bunch of packets with an FEC packet. A perfect analogy would be RAID (Redundant Array of Independent Disks), just like RAID, FEC packet contains parity of the packets in the FEC group. If any packet is lost then content of the packet can be recovered from the FEC packet and the remaining packets in the group. It is up to the sender to decide whether to send the FEC packets for optimizing specific scenarios or not.

\subsection{\large Connection Migration}
As mentioned in section \ref{tcp_con_mig}, TCP identifies flow by 5 tuple, connections in QUIC are being identified by a randomly generated 64 bit identifier called the \texttt{Connection ID}. In TCP, changing any parameter in 5-tuple can cause a connection break down and the session is no longer active. QUIC on the other hand, keep the \texttt{Connection ID} same and even if any parameter changes in 5-tuple underneath, it will not tear down the connection.

\subsection{\large Built-in Security}
In TCP world, choice of TLS depends on required security. It is an additional header gets imposed on top of TCP to encrypt the packet. This additional encryption comes with a cost of an extra handshake called TLS handshake. QUIC combines this handshake as a part of it's own handshake and do authentication of the end-points as well as negotiate all cryptography parameters. QUIC replaces TLS record layer with its own framing format, while keeping the same TLS handshake message. This ensures that connection is always authenticated and encrypted, but also in reduced RTTs. 

In addition to that QUIC also encrypts additional metadata so that middleboxes such as firewalls, proxies do not manipulate the connection.

\subsection{\large Solves Head-of-Line Blocking}
QUIC uses UDP as a transport layer protocol, which doesn't make packets hostage before sending them to application. Network stack sees QUIC packets as datagrams and doesn't require to reorder them. QUIC receives those datagrams and if implementation allows, can be pushed to application in any order necessary. This makes sure that packets are received as soon as network stack pushes them to the application.
QUIC also supports multiplexing which makes these packets in the same UDP pipeline and once it is established, no additional connection establishment overhead is required.

\begin{figure}[H]
    \centering
    \includegraphics[width=0.5 \linewidth]{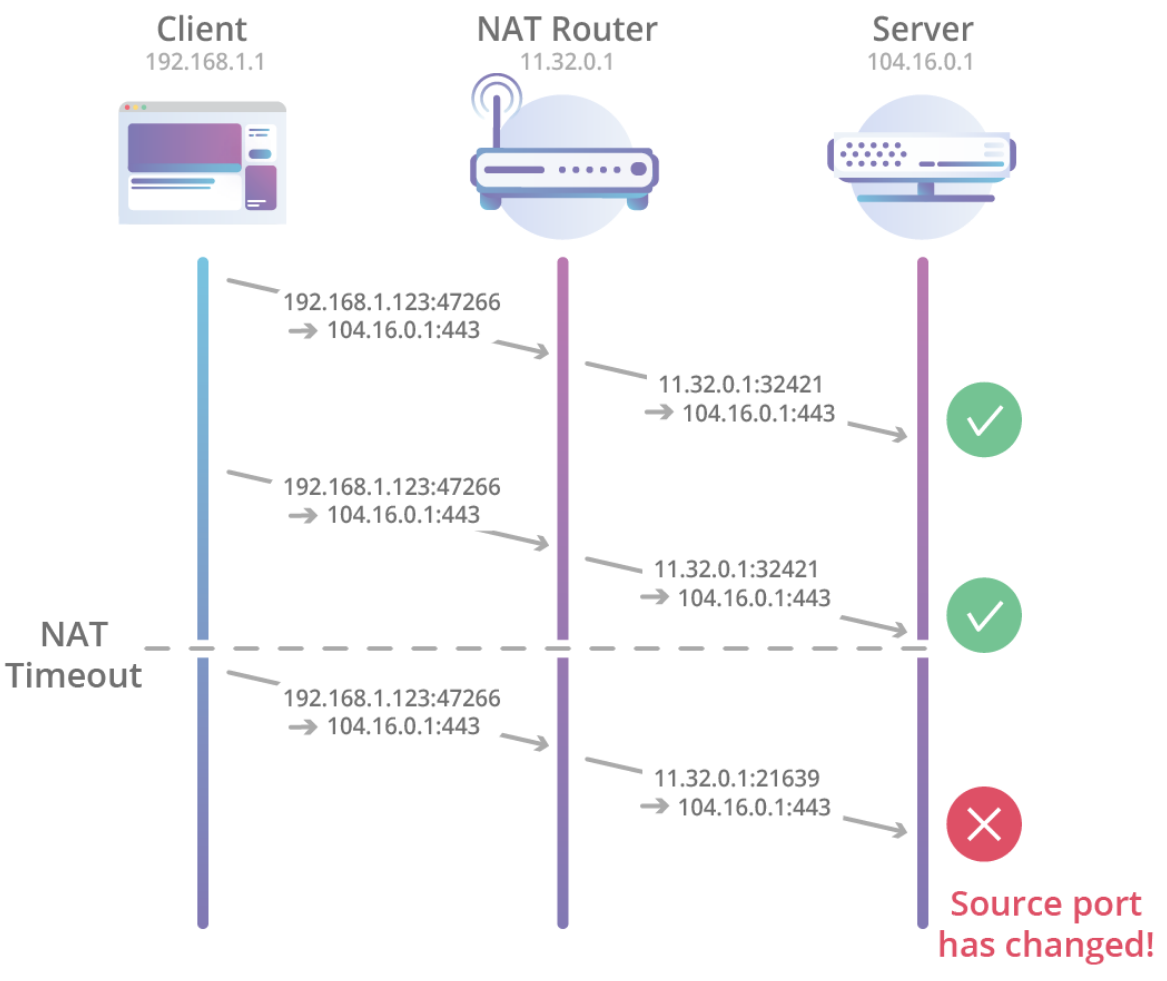}
    \caption{NAT Rebinding}
    \label{fig:NAT_rebinding}
\end{figure}

\subsection{\large NAT Rebinding}
Typical NAT routers have a NAT table to keep track of all the TCP connections passing through them. The hash to identify the connection is calculated via 5-tuple and by observing SYN, ACK and FIN packets. It is easy to detect when a new connection is established, which it is done by looking at the TCP table (TCP state table). 
Having UDP as a transport protocol, NAT routers cannot have the state of  a connection. NAT Routers will see this as UDP flows. When a NAT rebinding occurs, for instance, due to timeout, devices outside of the NAT perimeter will see those packets coming from different source port or IP than the one this flow was observed when the connection was originally established. This makes it impossible to track connections by only using the 5-tuple. Luckily, QUIC has connection ID which can be treated as hash of 5-tuple to identify the connection and the state can be determine by QUIC header (Described later).

\subsection{\large QPACK}
HTTP2 introduced header compression (HPACK)\cite{peon2015hpack}, which allows HTTP2 end hosts to reduce the amount of data transmitted. In particular HPACK has dynamic tables, filled by headers sent from previous HTTP requests/response. This allows end hosts to refer previously encountered headers in new requests, instead of having them transmitted again. HPACK tables need to be synchronized between HTTP sender and receiver. 
With TCP, synchronization is transparent, as TCP takes care of delivering HTTP requests and responses in the same order as they came in. On the other hand, QUIC can deliver multiple HTTP requests/responses, with the help of streams. This means that although it takes care of the data delivery in order, as long as it is a single stream, it doesn't guarantees any ordering among streams.

\section{\large QUIC Packet Format}
\begin{figure}[H]
    \centering
    \includegraphics[width=0.8 \linewidth]{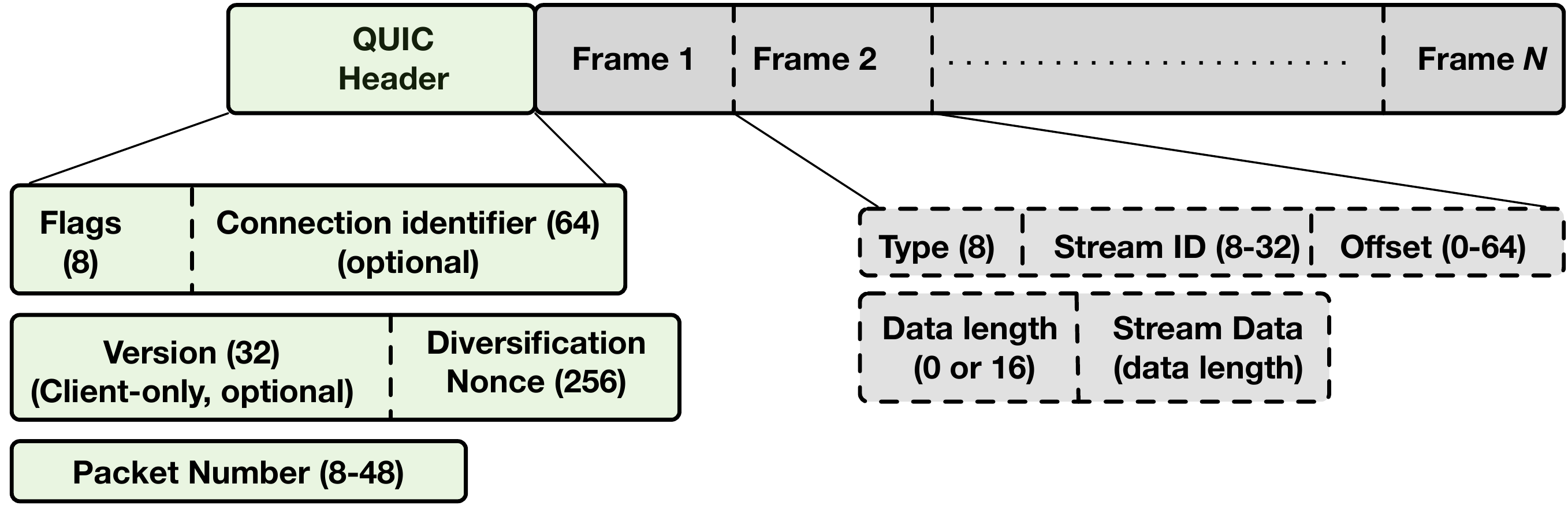}
    \caption{The solid and dashed lines show the clear-text and encrypted parts of a QUIC packet, respectively. The non-encrypted part is used for routing and decrypting the encrypted part of the packet.}
    \label{fig:packet_structure}
\end{figure}

Figure \ref{fig:packet_structure} shows the general packet format for QUIC. Generally, there are a lot of variants of those packet formats, such as Long header, short header, version negotiation packet, initial packet, 1-RTT packet, 0-RTT packet etc. Discussing those packet format is out of scope for this project, but interested readers can go \cite{QUICRfc} to read about those packet structures.

\subsection{\large QUIC Header}
QUIC header is of two types: long header and short header. As name suggests, long header is big in size and short header is small. Long header contains more information and have the following fields:
\begin{itemize}
    \item Version
    \item Destination Connection ID
    \item Source Connection ID
    \item Flags, such as header form, fixed bit, long packet type etc.
\end{itemize}

Table \ref{long_header_types} shows the types of long header packets in QUIC. Long header are the packets sent prior to the establishment of 1-RTT keys (Initial packets). Once the keys are negotiated and connection establishment is completed, sender switches back to short header. Long header allows special packets such as version negotiation packets etc to be represented in uniform fixed-length packet format. 
On the other hand short packets are the most seen packets in QUIC traffic. Short header packets are used after the version and 1RTT keys negotiation. Short header packet contains the following:

\begin{itemize}
    \item Destination Connection ID
    \item Packet Number
    \item Protected Payload
\end{itemize}

\begin{table*}[htp]
\centering

\begin{tabular}{|c|c|}
\toprule

\textbf{Type} & \textbf{Name} \\ \hline
0x0           & Initial       \\ \hline
0x1           & 0-RTT         \\ \hline
0x2           & Handshake     \\ \hline
0x3           & Retry         \\ \hline
\end{tabular}
\caption{Type of Long Header}
\label{long_header_types}
\end{table*}

\subsection{\large Connection Identifier}
Connection identifier is known as Connection ID or \texttt{CID}. Each connection has set of \texttt{CID}s, each of which can be used to distinguish a connection. \texttt{CID}s are independently selected by endpoints. The main task for \texttt{CID}s is to make sure that upon changing any lower payer protocols (UDP, IP) parameters, doesn't cause packets for a QUIC connection to be delivered to the wrong endpoint. Selecting \texttt{CID} is implementation specific, it is important because this method allows packets of that \texttt{CID} to be routed back to the endpoint and identified by the endpoint upon receipt.
At any given point, \texttt{CID} should not contain any information that can cause the middleboxes to co-relate them with other \texttt{CID}s for the same connection. In other words, \texttt{CID} must not be issued again for a single connection. 
A zero-length \texttt{CID} can be used if a \texttt{CID} is not needed to route to the correct endpoint. Although, multiplexing connections on the same local IP address and port while using zero-length \texttt{CID} will cause failures in the presence of peer connection migration, NAT rebinding, and client port reuse. Hence, must not be done unless an endpoint is certain that those protocol features are not in use.

\subsection{\large Stream Frame}
As we have discussed before, QUIC is a multiplexed protocol. All streams are multiplexed over one UDP connection. Stream Frames implicitly creates stream and carry data. Each QUIC stream has a flow control and lost data transmitted mechanism. Which means that if data is lost on stream then it does not effect others.

Stream frames consists of the following:
\begin{itemize}
    \item \textbf{\large Stream ID}: A variable length integer indicating the stream ID of the stream.
    \item \textbf{\large Offset}: A variable-length integer specifying the byte offset in the stream for the data in this stream frame. 
    \item \textbf{\large Length}: An integer of variable length. It specifies the length of the stream data field.
    \item \textbf{\large Stream Data}: Bytes to be delivered.
\end{itemize}

\subsection{\large Packet Number}
It is represented by an integer in the range of 0 to $2^{62-1}$. It is used in determining the cryptographic nonce of the packet for protection. Each end points maintains a separate packet number for sending and receiving. Packet numbers are limited to this range because they need to be represented in whole in the largest acknowledged field of ACK frame \cite{QUICRfc}. Packet numbers can be reduced and encoded in 1-4 bytes. Packet numbers are divided into 3 spaces in QUIC:
\begin{itemize}
    \item \textbf{\large Initial Space}: All initial packets are in this space.
    \item \textbf{\large Handshake Space}: All handshake packets are in this space.
    \item \textbf{\large Application Data Space}: All 0-RTT and 1-RTT encrypted packets are in this space.
\end{itemize}

A packet number space is a context in which a packet can be processed and acknowledged. For instance, initial packets can only be sent by initial packet protection keys. Similarly handshake packets are only sent at handshake encryption level. This brings cryptography separation between the data sent in the different packet sequence number spaces. Each space starts packets number as 0 and subsequent packets must increase the packet number by 1. 0-RTT and 1-RTT data exists in the same packet number space. There are certain additional rules to be followed in regards to packet number.

\begin{itemize}
    \item A QUIC end point must not reuse packet number within same packet number space in one connection.
    \item If a packet number reaches $2^{62-1}$, then the connection must be closed by sender without sending any notification. 
    \item A newly unprotected packet must be discarded by receiver, unless receiver is certain that another packet bearing same packet number has not been processed.
\end{itemize}

\begin{figure}[H]
    \centering
    \includegraphics[width=1 \linewidth]{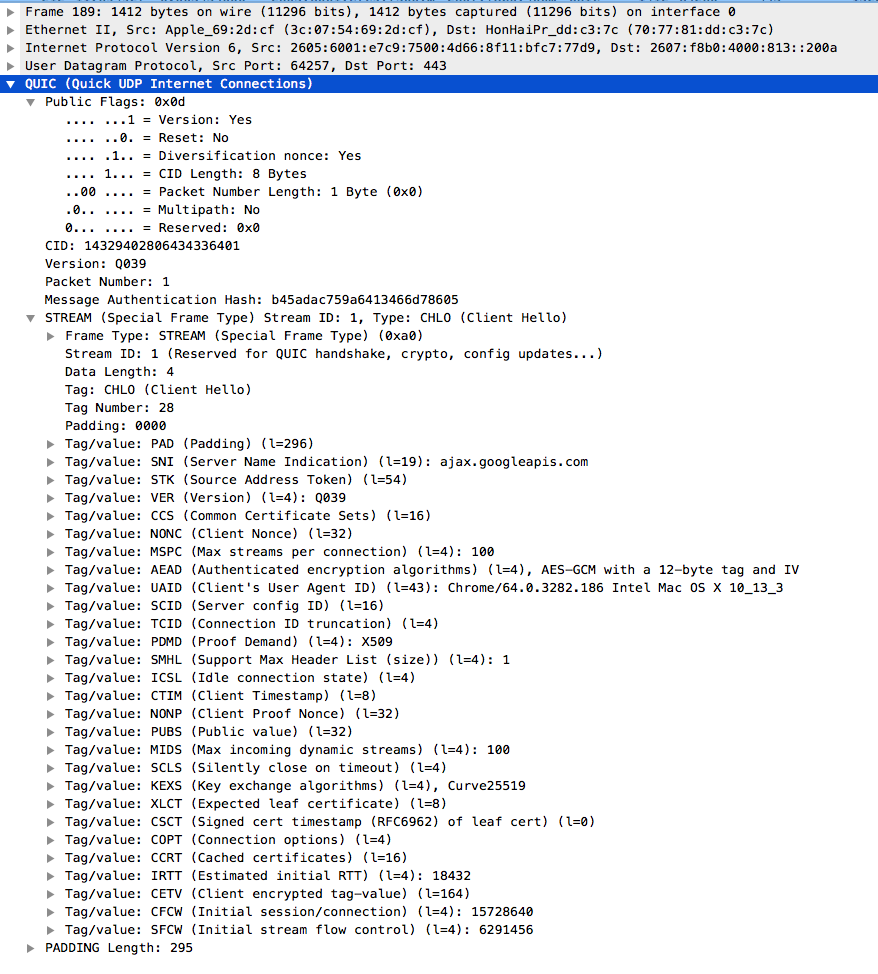}
    \caption{Inside of a QUIC Packet}
    \label{fig:packet_capture}
\end{figure}

\subsection{\large Flow Control}
As any connection oriented protocol demands, there has to be a way to limit the amount of data a receiver could buffer. In order to impose that limit, flow control is necessary. TCP implements it's own way of flow control \cite{speight2012method}. QUIC imposes flow control by controlling the maximum amount of data a sender can send on a stream at any given time. Also to limit concurrency within a connection, QUIC endpoint also controls the maximum cumulative number of streams that an endpoint can initiate. Since QUIC have TLS packets as a stream as shown in figure \ref{fig:packet_capture}. Those streams are referred as \texttt{CRYPTO} frames. They are not flow controlled like stream data because QUIC relies to have an interface to the TLS about the buffer limit.

\subsubsection{\large Data Flow Control}
QUIC data flow control is similar to credit based flow control in HTTP2. It means that receiver advertises the number of bytes it is ready to received on a given stream per connection. Which leads it to two level of data flow control:

\begin{itemize}
    \item \textbf{\large Stream Flow Control}: It prevents a single stream from consuming the entire buffer for a connection by limiting the amount of data that can be sent on any stream.
    \item \textbf{\large Connection Flow Control}: It prevents sender from exceeding a receiver's buffer capacity for the connection. It is done by limiting the total bytes of stream data sent in \texttt{STREAM} frames on all streams.
\end{itemize}

A receiver advertises credit for a stream by sending a \texttt{MAX\_STREAM\_DATA} frame with stream ID. \texttt{MAX\_STREAM\_DATA} frame indicates the maximum absolute byte offset of a stream. In addition to that, receiver also advertises credit for connection in \texttt{MAX\_DATA} frame. This frame indicates the maximum of the sum of the absolute byte offsets of all streams. To close connection, receiver uses \texttt{FLOW\_CONTROL\_ERROR} frame, if sender sends any corrupted packet, with regards to advertised connection or stream data limits.

Sender on the other hand, must ignore any \texttt{MAX\_STREAM\_DATA} or \texttt{MAX\_DATA} frames that do not increase any flow control limits. In case sender runs out of any credit, it will be unable to send any new data and will be considered blocked. In such event, a sender will send \texttt{STREAM\_DATA\_BLOCKED} or \texttt{DATA\_BLOCKED} frame to indicate that it has data to write but is blocked by flow control limits. 

\subsubsection{\large Flow Credits Increments}
An auto tuning mechanism is used by receiver to tune the frequency and amount of the advertised additional credit based on the RTT time estimate and the rate at which the receiving application consumes data, just like what TCP does. As mentioned earlier, if sender runs out of credit and unable to send new data, it is considered blocked. To avoid it, and to reasonably account for the possibility of data loss, a receiver should send a \texttt{MAX\_DATA} or \texttt{MAX\_STREAM\_DATA} frame at least two RTT before it expects the sender to get blocked. 
In addition, a receiver must not wait for a \texttt{STREAM\_DATA\_BLOCKED} or \texttt{DATA\_BLOCKED} frames before sending \texttt{MAX\_STREAM\_DATA} or \texttt{MAX\_DATA} frame, since doing so will mean that a sender will be blocked for at least an entire rount trip, and potentially longer if the peer chooses not to send \texttt{STREAM\_DATA\_BLOCKED} or \texttt{DATA\_BLOCKED} frame.

\subsection{\large Re-transmission}
In TCP world, packets are ACKed based on the sequence number and the next expected sequence number in the ACK packet. There are a lot of literature on TCP re-transmission, so we will not get into it. QUIC re-transmission is a bit different that TCP. As mentioned earlier, QUIC has multiple streams and any stream can be an ACK frame, acknowledging any past or present data frame. Due to this, QUIC maintains two counts, \texttt{ACK Range Count} and \texttt{First ACK Range}. \texttt{ACK Range Count} basically tells us till what packet number data frames has been received, and \texttt{First ACK Range} tells us till what packet number data frames have been received and acknowledged. 

\begin{figure}[H]
    \centering
    \includegraphics[width=1 \linewidth]{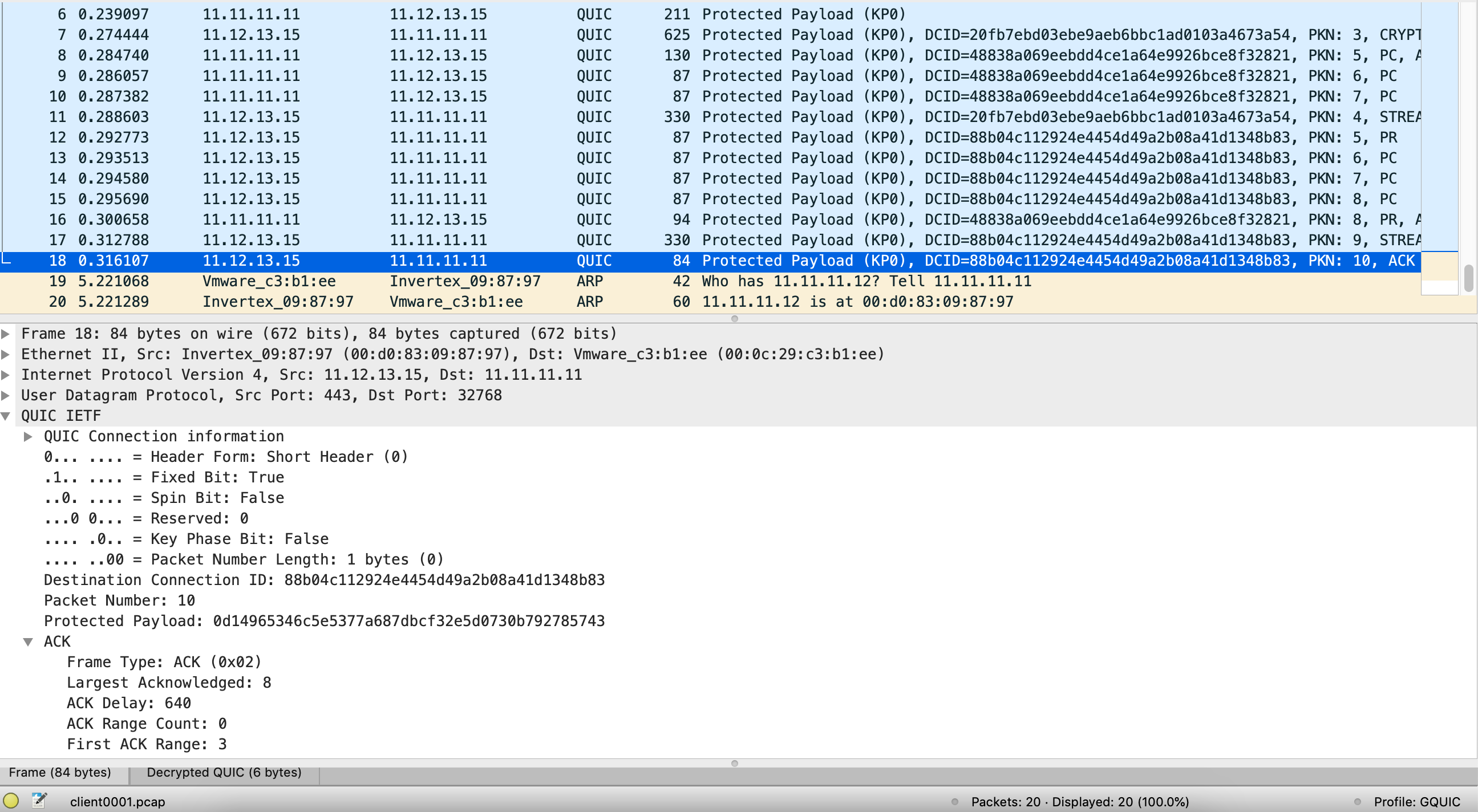}
    \caption{ACK Frame}
    \label{fig:ACK Frame}
\end{figure}

In case of any frame dropped, QUIC receiver tells sender via \texttt{Gap} in ACK frame. This tells sender how many packets were dropped and the NACK count tells what packet numbers are missing. Figure \ref{fig:ACK Frame} shows re-transmission ACK frame. 

\chapter{\large Conclusion and Future Aspects}
In this project, we examined different problems of TCP protocols faced today and explained their scenarios. In second chapter we dived down into QUIC protocol in detail. What we have found out is that QUIC protocol solves problems currently facing by TCP. Since QUIC sits in application layer, it brings substantial benefits, and it makes development and testing cycles fairly simple. 
QUIC is widely used in every type of traffic. Majorly it is used in user internet traffic and IoT traffic. We can conclude the QUIC protocol in these two parts. 

\subsection{\large User internet traffic}
Major goal of QUIC was an experimental transport protocol. It has now become a core part of a serving infrastructure. It was thought initially that wide deployment of a new UDP-based encrypted transport for HTTP was an audacious goal; there were many unknowns, including whether UDP blocking or throttling would be show-stoppers. In our QUIC protocol analysis we showed that QUIC overcome the shortcoming imposed in user internet traffic over TCP. QUIC provides layering in terms of streams, which squashes away the HTTP layers in QUIC, which allowed to remove inefficiencies in the HTTP's stack. Finally the tussle between end hosts are inevitable, it can only be resolved when all interested parties come to the table \cite{clark2002tussle}. Previously, attempts of deploying protocols that require any modification to network devices, such as SCTP \cite{maruyama2007stream}, TCP Fast open \cite{cheng2014tcp}-name a few, have unequivocally exposed the difficulties of incentives and achieving consensus on proposed network changes. 

\subsection{\large IoT Traffic}
IoT traffic is a bit different than user Internet traffic. Since IoT devices are usually deployed in lossy networks, the price they pay for a connection establishment while using TCP as a transport protocol is too much. QUIC mitigates that by introducing 0-RTT connection Establishment. QUIC eliminates the need of reconnecting upon changing any parameter in the 5-tuple. QUIC sees connection ID to identify a device and even if IP or port changes underneath, connection remains intact. QUIC solves head of line blocking problem by having streams mapped to one UDP pipeline. QUIC has better recovery mechanism by having FEC packets.

Table \ref{tcp_udp_quic_comp} shows the comparison between TCP, UDP, and QUIC.

\begin{table*}[htp]
\centering
\caption{Comparison of TCP, UDP and QUIC}
\begin{tabular}{|c|c|c|c|}
\hline
\textbf{Solves}                                                               & \textbf{TCP} & \textbf{UDP} & \textbf{QUIC} \\ \hline
Head-of-line Blocking                                                         & X            & N/A          & $\surd{}$             \\ \hline
Multiple Connection to server                                                 & $\surd{}$            & X            & $\surd{}$             \\ \hline
Connection Migration                                                          & X            & X            & $\surd{}$             \\ \hline
Connection Overhead                                                           & X            & X            & $\surd{}$             \\ \hline
Reliable Connection                                                           & $\surd{}$            & X            & $\surd{}$             \\ \hline
\begin{tabular}[c]{@{}c@{}}Flow Control and Congestion\\ Control\end{tabular} & $\surd{}$            & X            & $\surd{}$             \\ \hline
\end{tabular}
\label{tcp_udp_quic_comp}
\end{table*}

\subsection{\large Future Aspect}
Since QUIC sits in application layer, it utilizes CPU more than TCP. Modern NIC (Network Interface Card) has inbuilt TCP parameters offloading such as checksum and SSL offloading. QUIC is not there yet. CPU utilization in QUIC carries a heavy burden of encryption and decryption. QUIC also limits MTU (Maximum Transmission Unit) to 1392 bytes \cite{QUICRfc}, which means that fragmentation is not yet supported in QUIC.

\chapter{\large Acronyms}
Below are the majorly used acronyms.
\begin{table*}[htp]
\begin{tabular}{|c|c|c|}
\hline
\textbf{S.No.} & \textbf{Acronym} & \textbf{Full Form}                 \\ \hline
1              & TCP              & Transmission Control Protocol      \\ \hline
2.             & UDP              & User Datagram Protocol             \\ \hline
3.             & BIC              & Binary Increase Congestion Control \\ \hline
4.             & C2TCP            & Cellular Controlled Delay TCP      \\ \hline
5.             & NATCP            & Network Assisted TCP               \\ \hline
6.             & H-TCP            & Hamilton Institue-TCP              \\ \hline
7.             & HTTPS            & Hypertext Transfer Protocol Secure \\ \hline
8.             & HTTP             & Hypertext Transfer Protocol        \\ \hline
9.             & TLS              & Transport Layer Security           \\ \hline
10.            & RTT              & Round Trip Time                    \\ \hline
11.            & IP               & Internet Protocol                  \\ \hline
12             & ACK              & Acknowledgement                    \\ \hline
13.            & SSL              & Secure Socket Layer                \\ \hline
\end{tabular}
\end{table*}


\pagebreak

\nocite{*} 
\bibliographystyle{unsrt}
\bibliography{main} 
\addcontentsline{toc}{chapter}{\bibname}

\end{document}